\documentclass[10pt]{article}
\usepackage{amsmath,amssymb,amsfonts,amsthm,graphicx}
\title{Number-phase entropic uncertainty relations and Wigner functions for solvable quantum systems with discrete spectra}
\author{G.R. Honarasa, M. K. Tavassoly and M. Hatami
\\
\footnotesize{Atomic and Molecular Group, Faculty  of Physics, Yazd University, Yazd, Iran}
\\ \footnotesize{e-mail: honarasa@sutech.ac.ir, mktavassoly@yazduni.ac.ir  } }

\begin{document}

\maketitle \thispagestyle{empty}

 \begin{abstract}
  In this letter, the "number-phase entropic uncertainty relation" and the "number-phase Wigner function" of generalized coherent states associated to a few solvable quantum systems with non-degenerate spectra are studied. We also investigate time evolution of "number-phase entropic uncertainty" and "Wigner function" of the considered physical systems with the help of temporally stable Gazeau-Klauder coherent states.

 \end{abstract}

 {\bf keyword:}
   solvable quantum systems, nonlinear coherent states, entropic uncertainty relation, number-phase Wigner function

{\it PACS:} 42.50.Dv, 03.65.-w

\newcommand{\I}{\mathbb{I}}
\newcommand{\norm}[1]{\left\Vert#1\right\Vert}
\newcommand{\abs}[1]{\left\vert#1\right\vert}
\newcommand{\set}[1]{\left\{#1\right\}}
\newcommand{\R}{\mathbb R}
\newcommand{\C}{\mathbb C}
\newcommand{\DD}{\mathbb D}
\newcommand{\eps}{\varepsilon}
\newcommand{\To}{\longrightarrow}
\newcommand{\BX}{\mathbf{B}(X)}
\newcommand{\HH}{\mathfrak{H}}
\newcommand{\A}{\mathcal{A}}
\newcommand{\N}{\mathcal{N}}
\newcommand{\B}{\mathcal{B}}
\newcommand{\RR}{\mathcal{R}}
\newcommand{\HD}{\hat{\mathcal{H}}}
 \section{Introduction}\label{sec-intro}
 Quantum theory departs in many aspects from classical physics. One of these features is the well-known uncertainty principle. This principle plays an important role in the conceptual foundations of quantum mechanics and measurement theory.  There are two different mathematical formulations of the physical content of uncertainty relations, i) the Heisenberg uncertainty principle in terms of {\it "variances"} defined as $(\Delta \A)^2=\langle \A^2 \rangle -{\langle \A \rangle} ^2$ \cite{heisenberg} and ii) entropic uncertainty relation in terms of {\it "entropies"} of two canonically conjugate operators defined as $S_{\A}=-\sum_{n=0}^s P_a\ln P_a$ with $P_a$ as the probability distribution for each operator \cite{Bialynicki}. The main difference between the two formulations lies in the fact that entropic uncertainty relations only take the probabilities of the different outcomes of a measurement into account, while the in Heisenberg uncertainty principle the variances based on uncertainty relations depend also on the measured values (i.e., the eigenvalues of the observable) themselves (for a detailed discussion on the uncertainty relation for joint measurement refer to \cite{werner}). Recently, entropic uncertainty relation has gained an important role in the context of quantum optics \cite{guhne,ballester}. Entropic uncertainty relations for conjugate observables discussed in \cite{sanchez}. For many electromagnetic field states entropic uncertainty relations have been studied, for instance, Fock states and coherent states \cite{rojas}, binomial and negative binomial states \cite{joshi1}, squeezed states and Airy states \cite{joshi2} and multiphoton coherent states \cite{joshi}. \\
 On the other side, single mode nonlinear coherent states \cite{filho} or $f$-coherent states \cite{manko} have been constructed as right eigenstates of a deformed annihilation operator, $A|z,f\rangle=z|z,f\rangle$ where $A=af(n)$  with $a$ and $n$ as bosonic annihilation and number operator, respectively and so $f(n)$ is an intensity dependent function responsible for the nonlinearity of the states. The explicit form of these states in the number state bases are given by
 \begin{equation}\label{newket}
   |z,f\rangle=\mathcal{N}(\left|z\right|^2)^{-1/2}\sum_{n=0}^\infty d_n z^n|n\rangle,
 \end{equation}
 where $d_n\equiv\left(\sqrt{[nf^2(n)]!}\right)^{-1}$ with convention $d_0\dot{=}1$ and the normalization factor as $\mathcal{N}(\left|z\right|^2)=\sum_{n=0}^\infty d_n ^2 \left|z\right|^{2n}$. \\
 Very recently we have introduced the generalized coherent states corresponding to any quantum system with known discrete non-degenerate spectrum using nonlinear coherent states approach \cite{our,tavassoly} as,
 \begin{equation}\label{zen}
   |z,e_n\rangle =\mathcal{N}_e(\left|z\right|^2)^{-1/2}\sum_{n=0}^\infty \frac{z^n}{\sqrt{[e_n]!}}|n\rangle,
 \end{equation}
 where $[e_n]!\dot{=}\prod_{i=1}^n e_i$ with $[e_0]!\dot{=}1$. In (\ref{zen}) $e_n$ is the discrete spectrum of arbitrary physical system, i.e., $H|n\rangle=e_n|n\rangle$ with $e_0=0$ and $\mathcal{N}_e(\left|z\right|^2)$ is an appropriate normalization factor given by
 \begin{equation}\label{Ne}
   \mathcal{N}_e(\left|z\right|^2)=\sum_{n=0}^\infty \frac{\left|z\right|^{2n}}{[e_n]!}.
 \end{equation}

  Phase distribution and squeezing in number or phase operators of various physical systems with known discrete spectrum $e_n$ have been reported by us in \cite{our}. In the present work, we will further explore the relevance of the results obtained in \cite{our} to other popular information theoretical measure, namely the Shannon information entropy \cite{shannon}. In addition, the nonclassical properties of some classes of coherent states associated to a few solvable quantum systems with spectrum $e_n$ will be investigated  by studying the {\it "number-phase entropic uncertainty relations"}. We also pay attention to a special distribution function has been called {\it "number-phase Wigner function"} recently introduced in \cite{vaccaro}. We will reformulate these two quantities in terms of the spectrum of an arbitrary physical system.
  At last we consider Gazeau-Klauder coherent states as the dynamical evolution of the states in (\ref{zen}) and study the time evolution of number-phase entropic uncertainty relations and Wigner functions for these states.

 \section{Number-phase entropic uncertainty relations for generalized coherent states associated to solvable quantum systems}

 Let $\A$ and $\B$ be a pair of conjugate observables defined on an $(s+1)$-dimensional space, each with complete set of eigenstates $|a_n\rangle$ and $|b_n\rangle$ respectively, satisfying the eigenvalue equations
 \begin{equation}\label{eigenvalue2}
   \A|a_n\rangle=a_n|a_n\rangle,\;\;\;\; \B|b_n\rangle=b_n|b_n\rangle,
 \end{equation}
  where the discrete eigenvalues read as
 \begin{equation}\label{values}
   a_n=a_0+\frac{2\pi n}{(s+1)\beta},\;\;\;\; b_n=b_0+n\beta,
 \end{equation}
 with $a_0$, $b_0$ and $\beta$ as real constants. The normalization of the eigenstates and the requirement that $\A$ and $\B$ be canonically conjugate are respectively expressed by the relations \cite{pegg5}
 \begin{equation}\label{orth1}
   \langle a_n|a_m\rangle=\delta_{n,m}=\langle b_n|b_m\rangle,
 \end{equation}
 \begin{equation}\label{orth2}
   \langle a_n|b_m\rangle=\frac{\exp{(ia_n b_m)}}{\sqrt{s+1}}.
 \end{equation}
 If the probability that a measurement of $\A$ gives the result $a_n$ is denoted by $P_a(a_n)$, with similar expression for $\B$, then the {\it "Shannon entropies"} associated with the probability distributions for $\A$ and $\B$ were defined as \cite{shannon}
 \begin{equation}\label{sa}
   S_{\A}=-\sum_{n=0}^s P_a(a_n)\ln P_a(a_n),
 \end{equation}
 \begin{equation}\label{sb}
   S_{\B}=-\sum_{n=0}^s P_b(b_n)\ln P_b(b_n).
 \end{equation}
 These entropies are not independent but are bounded by an entropic uncertainty relation proposed by Kraus \cite{kraus} and has proven by Maassen and Uffink \cite{maassen}. According to their results for conjugate operators $\A$ and $\B$ in $(s+1)$-dimensional state space the following inequality holds
 \begin{equation}\label{sasb}
   S_{\A}+S_{\B}\geq \ln(s+1).
 \end{equation}
 Entropic uncertainty relation expressed in (\ref{sasb}) depends on dimension of state space and so for an infinite dimensional system this lower bound diverges. In order to make meaningful the statement in the infinite limit, Vaccaro et al. modified the definition of entropy and obtained a bound that is independent of dimension $s$ \cite{rojas}. Based on their proposal the differences $\delta_a=2\pi/[(s+1)\beta]$ and $\delta_b=\beta$, between successive eigenvalues of $\A$ and $\B$ respectively are constants and so they defined the new form of probability densities as
 \begin{equation}\label{papb}
   P_{\A}(n)=P_a(a_n)/\delta_a,\;\;\;\; P_{\B}(n)=P_b(b_n)/\delta_b.
 \end{equation}
 From the completeness and orthonormality of the eigenstates of $\A$ and $\B$ it follows that $\sum_{n=0}^s P_{\A}(n)\delta_a=1=\sum_{n=0}^s P_{\B}(n)\delta_b$. Now it is possible to define new quantities $R_{\A}$ and $R_{\B}$,
 \begin{equation}\label{ra}
   R_{\A}\equiv -\sum_{n=0}^s \delta_a P_{\A}(n)\ln P_{\A}(n)=S_{\A}+\ln \delta_a,
 \end{equation}
 \begin{equation}\label{rb}
   R_{\B}\equiv -\sum_{n=0}^s \delta_b P_{\B}(n)\ln P_{\B}(n)=S_{\B}+\ln \delta_b.
 \end{equation}
 Working with $R_{\A}$ and $R_{\B}$ instead of $S_{\A}$ and $S_{\B}$ one immediately finds a new form of entropic uncertainty relation which is independent of the dimension of space. Upon combining (\ref{sasb}), (\ref{ra}) and (\ref{rb}) we arrive at
 \begin{equation}\label{rarb}
   R_{\A}+R_{\B}\geq \ln(2\pi),
 \end{equation}
 where the equality holds for eigenstates of $\A$ and $\B$. The above inequality for the sum of the entropies of any pair of conjugate observables provides an alternative to Heisenberg uncertainty relation.\\
 The number and phase operator for a single mode electromagnetic field are conjugate operators.
 But searching for a hermitian phase operator of radiation field has a long history from the beginnings of quantum electrodynamics \cite{dirac}. There are several different approaches to the problem of quantum phase. Susskind and Glogower proposed an exponential phase operator \cite{susskind}. The idea was to perform a polar decomposition of the annihilation operator, similar to the polar decomposition of the complex amplitude performed for classical fields. Unfortunately, although their formalism permits to define associated hermitian operators, it has still a serious problem: it is non-unitary. Holevo introduced another approach \cite{holevo}. In this approach the mathematical representation of the concept of quantum observable is extended from a self-adjoint operator to a normalized positive operator measure (POM) and assume that any quantum phase observable is a phase shift covariant POM with the interval $[0,2\pi)$ as the range of its possible measurement outcomes. Later Barnett and Pegg \cite{barnet} introduced a unitary operator in an extended Hilbert space. The apparent difficulty of the latter proposal is that it included unphysical negative number states. Along these efforts, a very important development in this field has been made by Pegg and Barnett \cite{pegg,pegg2,pegg1}. They have defined a unitary and hermitian phase operator, and a phase state but in a finite although arbitrarily large subspace whose dimension was allowed to tend to infinity after the calculation of expectation values of observable quantities. Therefore, the Pegg-Barnett formalism has been extensively employed in recent literature on the phase properties of a wide variety of quantum systems in quantum optics \cite{vaccaro and others}. Moreover, the Pegg-Barnett formalism can also be embedded in the covariant approach and Lahti et al. showed that this formalism can be extended to cover the covariant theory \cite{lahti}.\\
 According to the Pegg-Barnett formalism a complete set of $(s+1)$ orthonormal phase states $|\theta_m\rangle$ are defined by \cite{pegg}
 \begin{equation}\label{tetaket}
   |\theta_m\rangle=\frac{1}{\sqrt{s+1}}\sum_{n=0}^s \exp{(in\theta_m)} |n\rangle,
 \end{equation}
 where $\left\{|n\rangle\right\}_{n=0}^s$ are the number states and $\theta_m$ is given by
 \begin{equation}\label{teta}
   \theta_m=\theta_0 +\frac{2\pi m}{s+1} ,\qquad   m=0,1,...,s,
 \end{equation}
 with arbitrary $\theta_0$ value. Based on the phase state definition in (\ref{tetaket}) the Hermitian phase operator is defined as \cite{roy}
 \begin{equation}\label{phi}
   \phi_\theta=\sum_{m=0}^s \theta_m |\theta_m\rangle \langle \theta_m|.
 \end{equation}
 Now, we identify our general conjugate operators $\A$ and $\B$ with the phase and number, $\phi_\theta$ and $n$, operators by choosing $\beta=1$, $a_0=\theta_0$ and $b_0=0$ in (\ref{values}). Henceforth, the entropic uncertainty relation for number and phase read as
 \begin{equation}\label{rphirn}
   R_\phi+R_n\geq \ln(2\pi).
 \end{equation}
 Now we are ready to consider the number-phase entropic uncertainty relations for generalized coherent states associated to solvable quantum
 systems. In the infinite $s$ limit the sum in (\ref{ra}) becomes a Riemann integral
 \begin{equation}\label{rphi}
   R_\phi=-\int_{\theta_0}^{\theta_0+2\pi} P(\theta)\ln P(\theta) d\theta,
 \end{equation}
 where $P(\theta)$ is the phase probability distribution of generalized coherent states associated to solvable quantum
 systems given by \cite{our}
 \begin{equation}\label{p}
   P(\theta)=\lim_{s\rightarrow\infty}\frac{s+1}{2\pi} \left|\langle\theta_m|z,e_n\rangle\right|^2.
 \end{equation}
 Taking into account the states (\ref{zen}) in (\ref{p}) one arrives at
 \begin{equation}\label{pteta3}
   P(\theta)=\frac{1}{2\pi}\left(1+2\mathcal{N}_e(|z|^2)^{-1}\sum_{n} \sum_{k<n} {\frac{z^n z^{*k}}{\sqrt{[e_n]![e_k]!}}}\cos{[(n-k)\theta]}\right)
 \end{equation}
 for the phase probability distribution. On the other side the entropy of the photon number is defined as
 \begin{equation}\label{rn0}
   R_n=-\sum_{n=0}^\infty \left|\langle n|z,e_n\rangle \right|^2 \ln{(\left|\langle n|z,e_n\rangle \right|^2)}
 \end{equation}
 where on using (\ref{zen}) can be straightforwardly inverted to
 \begin{equation}\label{rn}
   R_n=-\mathcal{N}_e(\left|z\right|^2)^{-1} \sum_{n=0}^\infty \frac{|z|^{2n}}{[e_n]!}\ln{\left(\mathcal{N}_e(\left|z\right|^2)^{-1}\frac{|z|^{2n}}{[e_n]!}\right)}.
 \end{equation}
 Note that $\mathcal{N}_e(\left|z\right|^2)$ in (\ref{pteta3}) and (\ref{rn}) has been defined in (\ref{Ne}).
    \section{The number-phase Wigner function of solvable quantum systems}
 The number-phase Wigner function for an arbitrary state is defined as the expectation value of the {\it "number-phase Wigner operator"} was expressed as \cite{vaccaro,vaccaro1}
 \begin{eqnarray}\label{wigner operator}
   \mathcal{W}_{np}(n,\theta)=\frac{1}{2\pi} \{\sum_{p=-n}^n e^{2ip\theta}|n+p\rangle \langle n-p| \nonumber \\ +\sum_{p=-n}^{n-1} e^{i(2p+1)\theta}|n+p\rangle \langle n-p-1| \},
 \end{eqnarray}
 where $n=0,1,2,3,...$ and $\theta$ is real. The second summation taken to be zero for $n=0$. There
 are remarkable similarities between this number-phase Wigner function and the usual position-momentum Wigner function in terms of their defining properties \cite{vaccaro}. In the latter formalism the problem of revivals found in the "discrete number-phase function" introduced by Vaccaro and Pegg \cite{vacc-pegg} do not appear and there is no need for fractional values of $n$ as used by Luks and Perinova \cite{luks}. So, it turns out that the number-phase Wigner function for generalized coherent states associated to solvable quantum systems may be defined as follows
 \begin{equation}\label{wigner f1}
   W_{np}(n,\theta)=\langle z,e_n|\mathcal{W}_{np}|z,e_n\rangle.
 \end{equation}
 Making use of the states (\ref{zen}) in (\ref{wigner f1}) one straightforwardly arrives at
 \begin{eqnarray}\label{wigner f2}
   W_{np}(n,\theta)=\frac{\mathcal{N}_e(\left|z\right|^2)^{-1}}{2\pi} \{ \sum_{p=-n}^n \frac {(z^*)^{n+p} z^{n-p} e^{2ip\theta}}{\sqrt{[e_{n+p}]![e_{n-p}]!}} \nonumber \\
 + \sum_{p=-n}^{n-1} \frac{(z^*)^{n+p} z^{n-p-1} e^{i(2p+1)\theta}}{\sqrt{[e_{n+p}]![e_{n-p-1}]!}}\} .
 \end{eqnarray}
 As we will observe in next section this distribution function may become negative in some regions which can be interpreted as the nonclassical signature of the physical states.
   \section{Physical applications of the formalism}
 Our results presented in sections 2 and 3 have the potentiality to be applied to arbitrary physical systems with known discrete spectra $e_n$'s. For this purpose we will pay attention to the following quantum systems.\\
  \emph{I)} {\it Hydrogen-like spectrum.} As an important physical system we will accomplish with is the hydrogen-like spectrum possesses the spectrum commonly described by \cite{tavassoly,Klauder}
 \begin{equation}\label{H-spect}
   e_n= 1- \frac {1}{(n+1)^2}\; .
  \end{equation}
 The associated coherent states may be obtained explicitly by setting (\ref{H-spect}) in (\ref{zen}) which are defined in the unit disk centered at the origin.\\
  \emph{II)} {\it P\"{o}schl-Teller potential.} The interest in this potential and its coherent states is due to various applications in many fields of physics particularly in atomic and molecular physics. This potential has the spectrum \cite{antoine}
 \begin{equation}\label{poschl}
   e_n= n(n+\nu),
 \end{equation}
 where $\nu\geq2$. The special case $\nu=2$ is specified to infinite square-well potential. The associated coherent states may be obtained explicitly by setting (\ref{poschl}) in (\ref{zen}) which are defined in the whole of complex plane.\\
  \emph{III)} {\it Isotonic oscillator.} As last example we will consider an interesting model of solvable systems has been called the isotonic oscillator Hamiltonian \cite{landau}. It is known that the corresponding Hamiltonian admits exact eigenvalues given by \cite{richard}
 \begin{equation}\label{isot}
   \epsilon_n= 2(2n+\gamma) , \qquad       n=0,1,2,...
 \end{equation}
 where $\gamma\geq \frac{3}{2}$. To satisfy the necessary condition $e_0=0$ so that $[e_0]!=1$ we have to use the shifted energy spectrum
 \begin{equation}\label{isot2}
   e_n= \epsilon_n-2\gamma=4n , \qquad       n=0,1,2,...\; .
 \end{equation}
 The associated coherent states may be obtained explicitly by setting (\ref{isot2}) in (\ref{zen}) which are defined in the whole of complex plane.

  We perform our numerical calculations based on the relations (\ref{rphi}) and (\ref{rn}) respectively for $R_\phi$, $R_n$ (and their sum), also (\ref{wigner f2}) for $W_{np}(n,\theta)$. Figs. 1, 2 and 3 show $R_\phi$, $R_n$ and their sum against $z\in \R$ for Hydrogen-like spectrum, P\"{o}schl-Teller potential and isotonic oscillator, respectively. As it may be expected in all cases the sum $R_\phi+R_n$ is $\ln{(2\pi)}$ at $z=0$ due to the fact that the vacuum is an eigenstate of number operator. In Fig. 1 $R_n$ and the sum of the entropies increase while $R_\phi$ decreases with increasing $z$. As it is seen from Figs. 2 and 3, after a few oscillations the sum of the entropies again tends to lower bound $\ln{(2\pi)}$ for large enough values of $z$. It can be observed that the lower bound of the sum $R_\phi+R_n$ associated to all of these states for all $z$ according to (\ref{rphirn}) is satisfied. In Figs. 4, 5 and 6 we have plotted the number-phase Wigner function for Hydrogen-like spectrum, P\"{o}schl-Teller potential and isotonic oscillator, respectively. The plots of the number-phase Wigner functions are represented in cylindrical coordinates as a surface at a height of $W_{np}(n,\theta)$ above the point $(x,y)$ with $x=n\cos{\theta}$ and $y=n\sin{\theta}$. The surface is drawn as curves of constant integer $n$ (concentric rings incremented by unit one) crossed by curves of constant $\theta$ (radial lines). In Fig. 4 the peak occurs on the $n=1$ curve, while in Figs. 5 and 6 the peaks are on the $n=3$ and $n=6$ curves, respectively. All peaks in the above three figures occur along $\theta=0$ direction. It is evident that in some regions the number-phase Wigner function is negative indicating the nonclassicality behavior of these states.

    \section{Gazeau-Klauder coherent states as the time evolution of the generalized coherent states associated to solvable quantum systems}
 At this point we would like noticing the link between Gazeau-Klauder coherent states and the states introduced in (\ref{zen}). Upon the proposal introduced by one of us in \cite{tavassoly}, by the action of the evolution type operator
 \begin{equation}\label{eval}
   S(\gamma)= e^{-i(\frac{\gamma}{\hbar \omega})H}
 \end{equation}
 on the states in (\ref{zen}) one can obtain the analytical form of Gazeau-Klauder coherent states as
 \begin{equation}\label{zgamma}
   |z,\gamma,e_n\rangle=\mathcal{N}_e(\left|z\right|^2)^{-1/2}\sum_{n=0}^\infty \frac{z^n e^{-ie_n\gamma}}{\sqrt{[e_n]!}}|n\rangle,
 \end{equation}
 where in (\ref{zgamma}) we have assumed $\hbar=1=\omega$ and $\gamma\in \R$. Setting $\gamma\equiv t$ in (\ref{eval}) one can interpret the states in (\ref{zgamma}) as the time evolution of the states in (\ref{zen}) \cite{tavassoly}. This concept provides us with a powerful and at the same time simple formalism to investigate the dynamical properties of {\it number-phase} "entropic uncertainty relation" and "Wigner function" as time goes on. Making use of (\ref{zgamma}) and following the same procedure lead us to (\ref{rn}), exactly the same result for $R_n$ is again obtained. But for the entropy associated with the phase operator the relation (\ref{rphi}) changes to
 \begin{equation}\label{rphigk}
   R_\phi(\gamma)=-\int_{\theta_0}^{\theta_0+2\pi} P(\theta,\gamma)\ln P(\theta,\gamma) d\theta,
 \end{equation}
 where $P(\theta,\gamma)$ is given by
  \begin{eqnarray}\label{ptetagk}
   P(\theta,\gamma)=\frac{1}{2\pi}\{1+2\mathcal{N}_e(|z|^2)^{-1}\sum_{n} \sum_{k<n} \frac{z^n z^{*k}}{\sqrt{[e_n]![e_k]!}} \nonumber \\ \times \cos{[(n-k)\theta]} \cos{[(e_n-e_k)\gamma]}\}.
 \end{eqnarray}
 Finally after using (\ref{zgamma}) in (\ref{wigner operator}) and (\ref{wigner f1}) the following expression for number-phase Wigner function can be obtained
 \begin{eqnarray}\label{wigner gk}
 W_{np}(n,\theta,\gamma)=\frac{\mathcal{N}_e(\left|z\right|^2)^{-1}}{2\pi} (\sum_{p=-n}^n \frac {(z^*)^{n+p} z^{n-p} e^{2ip\theta} e^{i(e_{n+p}-e_{n-p})\gamma}}{\sqrt{[e_{n+p}]![e_{n-p}]!}} \nonumber \\ +\sum_{p=-n}^{n-1} \frac{(z^*)^{n+p} z^{n-p-1} e^{i(2p+1)\theta}e^{i(e_{n+p}-e_{n-p-1})\gamma}}{\sqrt{[e_{n+p}]![e_{n-p-1}]!}}).
 \end{eqnarray}
 In Fig. 7a we have plotted the sum $R_n+R_\phi$ against real $z$ and $\gamma$ (dimensionless time) for hydrogen-like atoms. Fig. 7b is the cross section of Fig. 7a for various fixed values of $z\in \R$. The figure shows damping behavior of oscillation as $\gamma$ goes on. It is seen that as $z$ increases the amplitude of oscillations increases.
 We have plotted the sum $R_n+R_\phi$ against $z$ and $\gamma$ for P\"{o}schl-Teller potential in Fig. 8a. It's oscillatory behavior with $\gamma$ is clearly seen. In Fig. 8b we have displayed the cross section of Fig. 8a for some fixed values of $z\in \R$. It is observed that the number of oscillations and their amplitudes grow up rapidly with increasing $z$ values.
 In Fig. 9a we have plotted the sum $R_n+R_\phi$ against $z$ and $\gamma$ associated to isotonic oscillator. Interestingly it does not change with $\gamma$. This feature has been shown apparently in Fig. 9b which is a cross section of Fig. 9a for some different values of $z$. So that the sum of the number and phase entropies for isotonic oscillator is a constant quantity with time, i.e., it is conserved.

    \section{Summary and conclusion}
 In summary, we obtained a generalized formulation for entropic uncertainty relations of nonlinear coherent states and extended it to arbitrary solvable quantum systems with discrete non-degenerate spectra. We have examined the number-phase entropic uncertainty relations of generalized coherent states associated to a few well-known solvable quantum systems. It is found that for all of these states, number-phase entropic uncertainty relation have a common lower bound. In other words the inequality in (\ref{rphirn}) associated to all of the considered physical examples is satisfied. We have also studied the number-phase Wigner representation $W_{np}(n,\theta)$ for the considered physical examples. We have illustrated the advantage of $W_{np}(n,\theta)$ by giving a direct graphical representation of the number-phase which was not possible from Wigner's original probability distribution. To this end, the time evolution of the number-phase "entropic uncertainty relation" and "Wigner function" for generalized coherent states of the physical systems are discussed. As a remarkable point it is worth noticing that while for some of quantum physical systems the sum of the entropies of the number and phase operators oscillates with time, for the isotonic oscillator it is a conserved quantity.\\\\
\vspace {0.2 cm}
 {\bf Acknowledgements}\\
 The authors would like to thank the referees for their useful comments which allow to to improve the paper considerably.
   


 \newpage
 {\bf FIGURE CAPTIONS}

 \vspace {.5 cm}

  {\bf FIG. 1} Plot of $R_\phi$ (dot-dashed curve), $R_n$ (dashed curve) and their sum (solid curve) against $z\in \R$ associated to hydrogen-like spectrum.

 \vspace {.5 cm}

  {\bf FIG. 2} Plot of $R_\phi$ (dot-dashed curve), $R_n$ (dashed curve) and their sum (solid curve) against $z\in \R$, keeping $\nu$ fixed at 5 associated to P\"{o}schl-Teller potential.

 \vspace {.5 cm}

  {\bf FIG. 3} Plot of $R_\phi$ (dot-dashed curve), $R_n$ (dashed curve) and their sum (solid curve) against $z\in \R$, keeping $\gamma$ fixed at 5/2 associated to isotonic oscillator.
 \vspace {.5 cm}

  {\bf FIG. 4}  Plot of the number-phase Wigner function associated to hydrogen-like spectrum with $z=0.5$.

 \vspace {.5 cm}

  {\bf FIG. 5}  Plot of the number-phase Wigner function associated to P\"{o}schl-Teller potential with $z=5$.

 \vspace {.5 cm}

  {\bf FIG. 6} Plot of the number-phase Wigner function associated to isotonic oscillator with $z=5$.

 \vspace {.5 cm}

  {\bf FIG. 7}  (a) Three-dimensional plot of $R_\phi+R_n$ for hydrogen-like spectrum. (b) Two-dimensional plot of $R_\phi+R_n$ against $\gamma$ with $z=0.3$ (dot-dashed curve), $z=0.5$ (dashed curve) and $z=0.8$ (solid curve) for hydrogen-like spectrum.

 \vspace {.5 cm}

  {\bf FIG. 8}  (a) Three-dimensional plot of $R_\phi+R_n$ for P\"{o}schl-Teller potential. (b) Two-dimensional plot of $R_\phi+R_n$ against $\gamma$ with $z=2$ (dot-dashed curve), $z=5$ (dashed curve) and $z=20$ (solid curve) for P\"{o}schl-Teller potential.

 \vspace {.5 cm}

  {\bf FIG. 9}  (a) Three-dimensional plot of $R_\phi+R_n$ for isotonic oscillator. (b) Two-dimensional plot of $R_\phi+R_n$ against $\gamma$ with $z=2$ (dot-dashed curve), $z=10$ (dashed curve) and $z=20$ (solid curve) for isotonic oscillator.
 \end{document}